\documentclass[a4paper,prl,twocolumn,showpacs,superscriptaddress,floatfix]{revtex4-2}
\usepackage{graphicx}
\usepackage{dcolumn}
\usepackage{bm}
\usepackage{hyperref}
\hypersetup{
  colorlinks=true,        
  linkcolor=blue,         
  citecolor=cyan,         
  urlcolor=cyan            
}
\usepackage{colortbl}
\usepackage{amsmath}
\usepackage{times}
\usepackage{times}
\usepackage{multirow}
\begin{document}
\title{Inferring the nuclear symmetry energy at supra saturation density from neutrino cooling}

\author{Tuhin Malik}
\email{tuhin.malik@gmail.com}
\affiliation{CFisUC, Department of Physics, University of Coimbra, 3004-516 Coimbra, Portugal.}
\author{B. K. Agrawal}
\email{bijay.agrawal@saha.ac.in}
\affiliation{Saha Institute of Nuclear Physics, 1/AF Bidhannagar, Kolkata 700064, India.} 
\affiliation{Homi Bhabha National Institute, Anushakti Nagar, Mumbai 400094, India.} 
\author{Constan\c ca Provid\^encia}
\email{cp@uc.pt}
\affiliation{CFisUC, Department of Physics, University of Coimbra, 3004-516 Coimbra, Portugal.}

\date{\today}

\begin{abstract} 
An ambitious goal of the astrophysical community is not only  to constrain the equation of state (EOS) of neutron star (NS) matter by confronting it with astrophysics observations, but ultimately also to infer the NS composition. Nevertheless, the composition of the NS core is likely to remain uncertain unless we have an accurate determination of the nuclear symmetry energy at supra saturation density ($\rho>\rho_0$). We investigate how the nucleonic direct Urca  (dUrca) processes can be used as an effective probe to constraint the high density nuclear symmetry energy. A large  number of minimally constrained EOSs has been constructed by applying a Bayesian approach to study the correlations of the symmetry energy at different densities with a few selected properties of a NS. The nuclear symmetry energy above the baryon density 0.5 fm$^{-3}$ ($\sim 3 \rho_0$) is found to be strongly correlated with NS mass at which the onset of nucleonic dUrca neutrino cooling takes place in the core. This allows us to constrain the high density behavior of nuclear symmetry energy within narrow bounds. {The pure neutron matter pressure constraint from chiral effective field theory rules out the onset of nucleonic dUrca in stars with a mass $\lesssim$ 1.4 $M_\odot$.} The onset of dUrca inside 1.6 M$_\odot$ to  1.8 M$_\odot$ NS implies a  slope of the symmetry energy $L$ at $\sim 2.5~\rho_0$, respectively, between 54 and 48 MeV. 
\end{abstract}

\keywords{Nuclear Symmetry Energy -- Neutron star -- Nucleonic direct Urca}  

\maketitle
{\it Introduction.} Neutron Stars (NS) still remain one of the most intriguing astrophysical objects. Their extreme conditions far beyond the ones reachable in terrestrial laboratories, e.g., the very large neutron-proton asymmetry and baryonic density, make NS unique systems to understand the physics of matter at extreme conditions. The composition of their core still remains uncertain, but this information is essential to know the behavior of matter at such extreme conditions. Recent developments in multi-messenger astronomy bring important information on high-density nuclear matter physics. The values of the tidal deformability extracted from gravitational wave (GW) events, such as the GW170817 \cite{LIGOScientific:2018cki,LIGOScientific:2018hze} associated to a binary NS merger,  or the simultaneous measurement of NS masses and radii from the high-precision X-ray space missions, such as the NICER (Neutron star Interior Composition ExploreR) which recently measured the mass and radius of the pulsars PSR J0030+0451 \cite{Riley:2019yda,Miller:2019cac} from NICER and PSR J0740+6620 \cite{Riley:2021pdl,Miller:2021qha} from NICER and XMM-Newton Data, have already shed some light on the equation of state (EOS) of neutron star matter, which is a key quantity that connects the microscopic physics of NS matter to their macroscopic properties. The ambitious goal of the astrophysical community is not only to narrow down the {uncertainties in the } EOS, i.e., the possible scenarios for NS matter, by confronting them with astrophysics observations, but ultimately also to infer the different particle species present in the stars. However, the interior composition of the NS core is likely to remain uncertain even if one assumes that only nucleons, together with electrons and muons, are present in the neutron star core. The extraction of the nuclear matter properties from the $\beta$-equilibrium EOS has proven to be impossible without the knowledge of the composition or symmetry energy at high densities \cite{Tovar2021,Imam2021,Mondal2021} or the knowledge of the EOS of symmetric nuclear matter \cite{Essick2021}.

The composition inside NS is fixed by the weak-interaction, i.e, the $\beta$ equilibrium condition. The interior chemical composition has a direct effect on the NS cooling. The NS cooling, both in middle-aged isolated neutron stars and in  accreting neutron stars, is largely affected by the presence or absence of nucleon dUrca neutrino cooling in the star’s core. This process is about $10^6$ times more efficient than the modified Urca process \cite{Yakovlev:2000jp,Yakovlev:2004iq}. The direct Urca neutrino cooling, which involves the conversion of neutrons into protons and vice-versa via the weak interaction, requires a minimum proton fraction of 1/9 \cite{Lattimer:1991ib}, and a slightly  larger fraction after the onset of muons. As the proton fraction  $y_p$ increases with the baryonic density, dUrca processes may {occur} inside massive NSs if $y_p$ crosses the critical proton fraction threshold.

It has been shown in previous studies \cite{Klahn:2006ir,Cavagnoli:2011ft,Providencia:2013dsa,Fortin:2016hny,Providencia:2018ywl}  that the dUrca process is strongly influenced by the density dependence of the symmetry energy because this quantity controls the proton fraction inside the NS. In \cite{Klahn:2006ir},  it was discussed that  some models predict the opening of the dUrca at too low masses, and it was considered reasonable that  the onset of dUrca processes  occurs at masses $\gtrsim 1.5\, M_\odot$.   According to a recent study which has applied a statistical approach to describe the thermal evolution isolated NS and accreting NS \cite{Beznogov:2015ewa},  a successful description of the cooling curves is obtained considering that the opening of the dUrca occurs in stars with masses $\approx 1.6-1.8\, M_\odot$. {The existence of a clear anti-correlation between the onset of dUrca and the slope of the symmetry energy $L$  at saturation was shown in \cite{Cavagnoli:2011ft,Fortin:2016hny,Alam2015}. These calculations were performed using relativistic mean field (RMF) models with constant couplings. However,  for RMF models with nucleon-meson density dependent couplings {referred to as DDH models in the following)}  such as DD2 and DDME2 \cite{Fortin:2016hny} or the complete set of models generated in \cite{Malik:2022zol}, the nucleonic dUrca process was not found to occur inside NS.}

In the present communication we look into the {question "How can  the proton fraction inside the NS core be constrained?"}. It is well understood that the precise and simultaneous measurements of NS properties such as mass, radius, moment of inertia and tidal deformability may constrain the NS matter EOS within a narrow range \cite{Lindblom2012,Ozel:2010fw,Steiner:2010fz,Steiner:2012xt,Raithel:2016bux,Fujimoto:2017cdo,Fujimoto:2019hxv}. In order to further decompose a NS matter EOS to its isoscalar and isovector part, we need to know the symmetry energy at high density. {We will use the close relation between the onset of the nucleonic dUrca and the proton fraction to answer this question.}

{It will be shown that {DDH} RMF models do not predict nucleonic dUrca because of the density dependence of the $\rho$-meson coupling. This coupling  decreases at high densities to very small values, and, therefore, favors very asymmetric matter. {Although, this behavior of the $\rho$-meson coupling does not affect much the high density EOS of stellar matter, it has a strong effect on the composition, e.g. the charge fraction or electron and neutrino content.} In the present study, we propose a different parametrization of the  $\rho$-meson coupling that avoids this restriction at high densities, and consider the {threshold density} of the nucleonic  dUrca process as a probe to constrain its high density behavior. The NS mass with this threshold density at the center will be referred here after  M$_{\rm dUrca}$.}

A large set of minimally constrained EOSs, corresponding to the  DDH relativistic nuclear models is constructed  via a Bayesian inference  approach. We further extend our analysis, and employ  additional pseudo data on the M$_{\rm dUrca}$ to  constrain the  high density behavior of  symmetry energy. It will be shown that for densities above 0.5 fm$^{-3}$, the symmetry energy is linearly correlated  with M$_{\rm dUrca}$, and the coefficients of the linear regressions are parameterized as a function of the baryon density {with an maximum uncertainty  $\sim 6\%$.} 

{\it DDH Framework.}
The interactions among nucleons can be modeled within a RMF framework with an effective Lagrangian involving baryon and meson fields: the force between two nucleons is realized by the exchange of mesons. The $\sigma$ meson creates a strong attractive central force,  the $\omega$-meson is responsible for the repulsive short range force, and both determine the spin-orbit potential. The isovector $\varrho$ meson is included to distinguish between neutrons and protons, and introduce the isospin symmetry and  independence of the nuclear force. The Lagrangian including the nucleon field, the $\sigma$, $\omega$ and $\varrho$ mesons and their interactions can
be written as,
\begin{equation}
\begin{aligned}
\mathcal{L}=& \bar{\Psi}\Big[\gamma^{\mu}\left(i \partial_{\mu}-\Gamma_{\omega} A_{\mu}^{(\omega)}-
\Gamma_{\varrho} {\boldsymbol{\tau}} \cdot \boldsymbol{A}_{\mu}^{(\varrho)}\right) \\
&-\left(m-\Gamma_{\sigma} \phi\right)\Big] \Psi 
+ \frac{1}{2}\Big\{\partial_{\mu} \phi \partial^{\mu} \phi-m_{\sigma}^{2} \phi^{2} \Big\} \\
&-\frac{1}{4} F_{\mu \nu}^{(\omega)} F^{(\omega) \mu \nu} 
+\frac{1}{2}m_{\omega}^{2} A_{\mu}^{(\omega)} A^{(\omega) \mu} \\
&-\frac{1}{4} \boldsymbol{F}_{\mu \nu}^{(\varrho)} \cdot \boldsymbol{F}^{(\varrho) \mu \nu} 
+ \frac{1}{2} m_{\varrho}^{2} \boldsymbol{A}_{\mu}^{(\varrho)} \cdot \boldsymbol{A}^{(\varrho) \mu},
\end{aligned}
\label{lagrangian}
\end{equation}
where  $\Psi$ is the Dirac spinor for spin $\frac{1}{2}$ particles, and, in the present calculation, describes a nucleon doublet (neutron and proton) with bare mass $m$. The $\gamma^\mu $ and $\boldsymbol{\tau}$ are the Dirac matrices and the Pauli matrices, respectively. The vector meson field strength tensors are given by   $F^{(\omega, \varrho)\mu \nu} = \partial^ \mu A^{(\omega, \varrho)\nu} -\partial^ \nu A^{(\omega, \varrho) \mu}$. The $\Gamma_{\sigma}$, $\Gamma_{\omega}$ and $\Gamma_{\varrho}$ are the coupling constants of nucleons to the meson fields $\sigma$, $\omega$ and $\varrho$, respectively, and the corresponding meson masses are $m_\sigma$, $m_\omega$ and $m_\varrho$. A DDH model is considered with nucleon-meson   density-dependent coupling parameters in the form of
\begin{equation}
  \Gamma_{M}(\rho) =\Gamma_{M,0} ~ h_M(x)~,\quad x = \rho/\rho_0~,
\end{equation}
where the  density $\rho$ is the baryonic density, the $\Gamma_{M,0}$ is the coupling at saturation density $\rho_0$ and $M \in \{ \sigma, \omega, \varrho \}$. For the isoscalar couplings,  in the present study  the function $h_M$ is given by
\begin{equation}
h_M(x) = \exp[-(x^{a_M}-1)], \quad M=\sigma, \, \omega
\label{hm1}
\end{equation}
and  the isovector coupling is defined as
\begin{equation}
h_\varrho(x) = y ~ \exp[-a_\varrho (x-1)] + (y-1) ~, \quad 0<y\le 1~,
\label{hm2}
\end{equation}
{being a generalization of the  form proposed in \cite{Typel1999}. The parametrization defined in Eq. (\ref{hm1}) introduces  for the isoscalar meson couplings only  one extra parameter,  besides $\Gamma_{M,0}$.  The density  dependence of these couplings reproduce the  behavior obtained within a Dirac-Br\"uckner-Hartree-Fock calculation \cite{TerHaar1987,Brockmann1990,Typel1999}, for $\rho\gtrsim 0.04$ fm$^{-3}$. This range of densities is adequate to describe the NS core EOS. For the $\varrho$-meson coupling, two extra parameters are introduced with $y$ controlling the high density coupling value.
 The M$_{\rm dUrca}$ allows to constrain this parameter.}

\begin{figure}
\centering
\includegraphics[width=0.48\textwidth,angle=0]{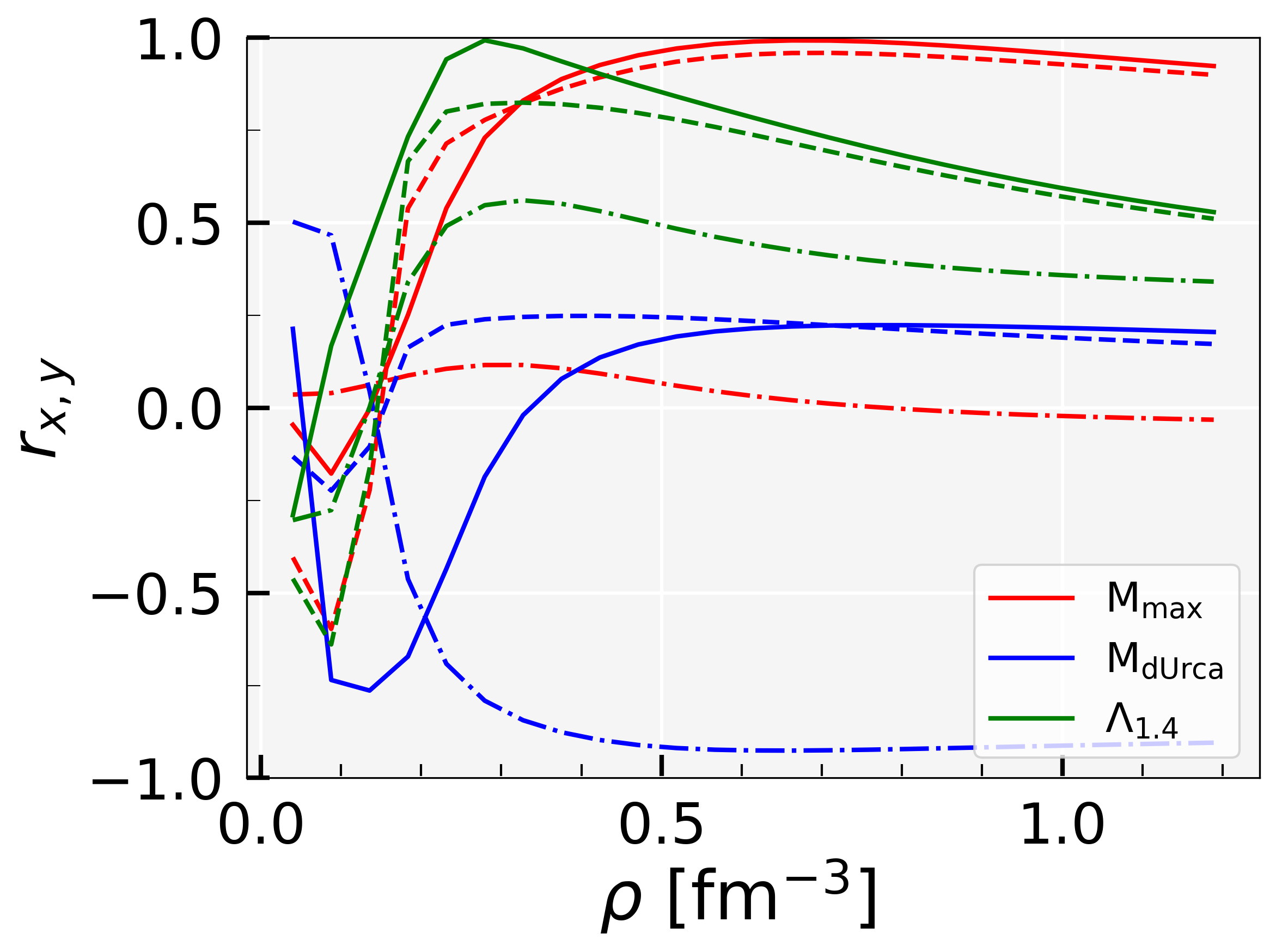}
\caption{The correlation coefficient $r_{x,y}$ as a function of the baryon density for $x$ the  $\beta-$equilibrium pressure (full lines),  the  SNM pressure (dashed lines) and the nuclear symmetry energy (dot-dashed lines) with $y$ equal to the NS maximum mass M$_{\rm max}$ (red),  the M$_{\rm dUrca}$ (blue)  and the dimensionless tidal deformability $\Lambda_{1.4}$ for 1.4 M$_{\odot}$ NS (green). \label{cor}}
\end{figure}

\begin{figure*}[t]
\begin{tabular}{lr}
\includegraphics[width=0.62\textwidth,angle=0]{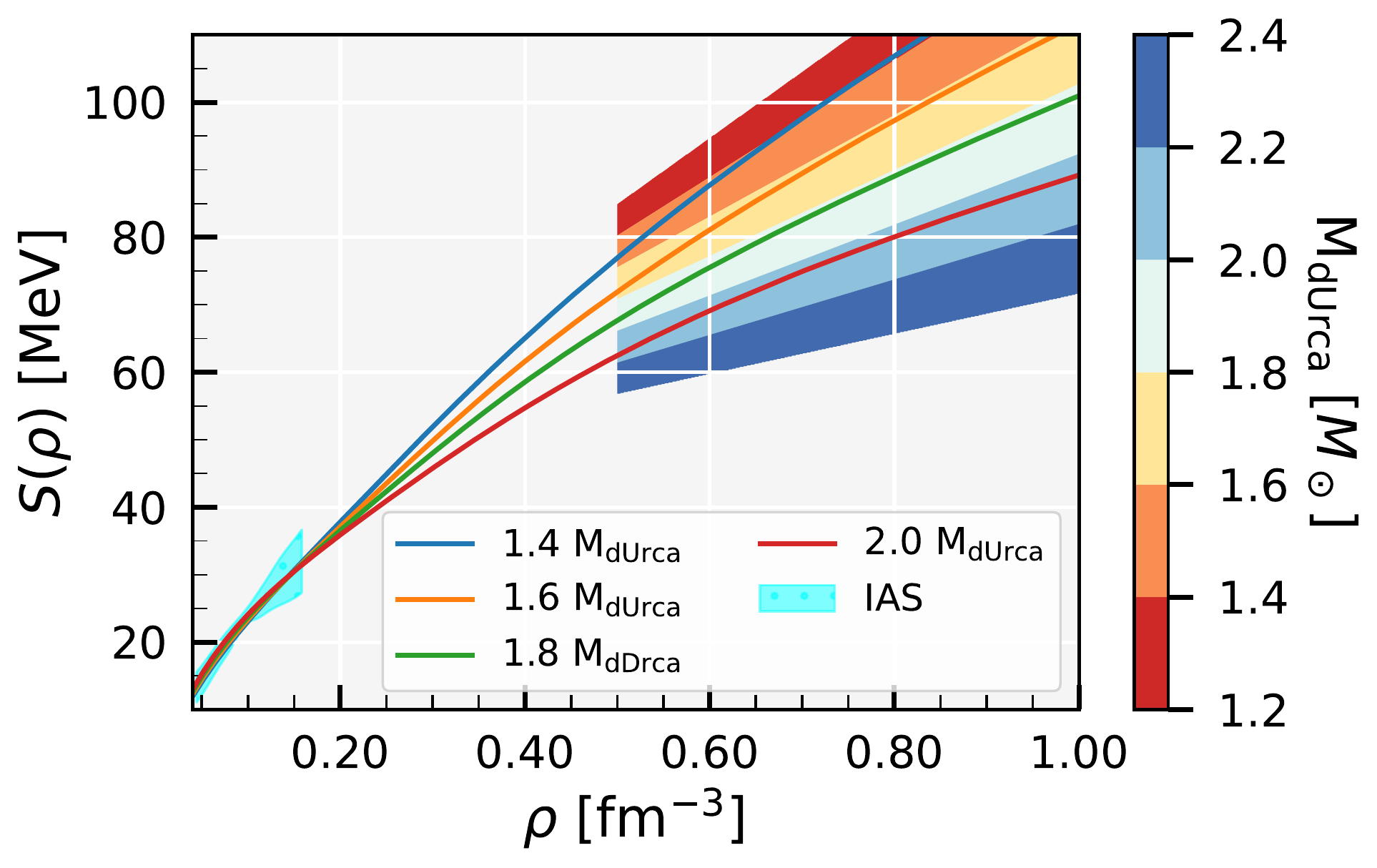}&
\includegraphics[width=0.36\textwidth,angle=0]{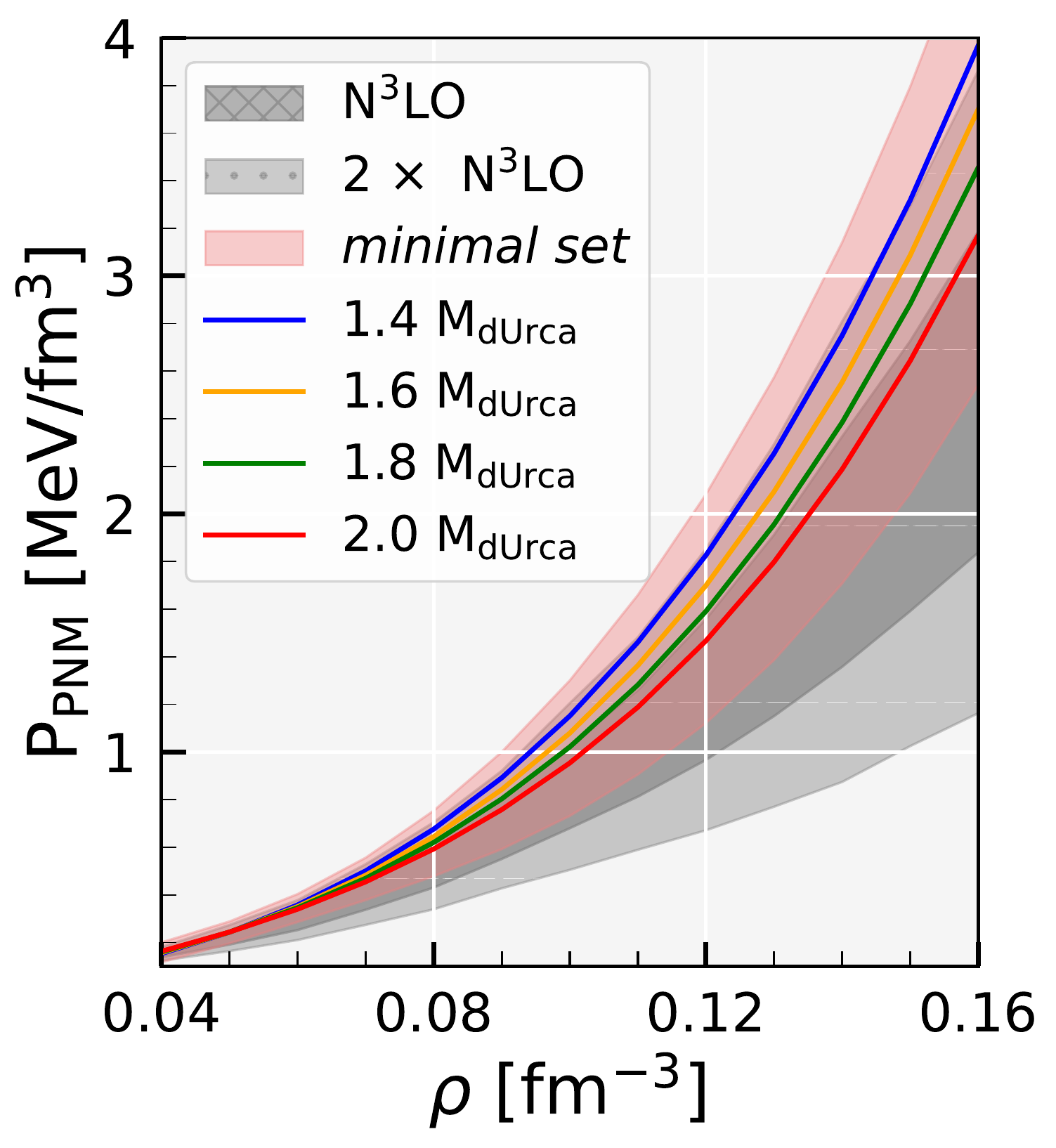}
\end{tabular}
\caption{Left: The density dependence of the nuclear symmetry energy $S(\rho)$ for  median values obtained with the {\it minimal set} and the additional dUrca mass constraint imposed for 1.4, 1.6, 1.8 and 2.0 M$_\odot$ NS (respectively the blue, orange, green and red  lines). The color bands were obtained from  the correlation relation of symmetry energy with the dUrca mass, Eq. (\ref{srho}),  considering the  EOS set that satisfies the {\it minimal set}. The constraints from the IAS \cite{Danielewicz:2013upa} bands are also shown. Right: The pressure of low density neutron matter from a N$^3$LO calculation in $\chi$EFT \citet{Hebeler2013} [with 1 (2) $\sigma$ uncertainty dark (light) gray], the 90\% CI of low density pure neutron matter pressure obtained with the {\it minimal set}, and the median lines for the sets with the additional dUrca mass constraint imposed for 1.4, 1.6, 1.8 and 2.0 M$_\odot$ NS.
\label{esym}}
\end{figure*}

{\it Bayesian Framework. \label{bayes}}
We apply a Bayesian approach to construct a large number of minimally constrained EOSs, along with  their individual  isoscalar and isovector components, and study the correlations between the EOS properties and  a few selected star properties, including the M$_{\rm dUrca}$. In this framework, the model parameters of DDH are informed by a given set of minimal fit data.  The posterior distributions of the model parameters $\theta$ in Bayes’ theorem can be written as  
\begin{equation}
P(\bm{\theta} |D ) =\frac{{\mathcal L } (D|\bm{\theta}) P(\bm {\theta })}{\mathcal Z},\label{eq:bt}
\end{equation}
where  $D$ denotes the set of  fit data,  $P(\bm {\theta })$ is the prior for the  model parameters and $\mathcal Z$ is the evidence.

The fit data considered are (i) a base set of fit data, referred to hereafter as "{\it minimal set}", and (ii) four additional  pseudo data for the M$_{\rm dUrca}$ identifying four different scenarios, and read as
 follows:
\begin{itemize}
    \item  The {\it minimal set} includes the nuclear saturation density $\rho_0=0.153\pm$0.005 fm$^{-3}$, the binding energy per nucleon $\epsilon_0=-16.1\pm0.2$, {the} incompressibility coefficient $K_0=230\pm40$, the symmetry energy $J_{\rm sym,0}=32.5\pm1.8$, all evaluated at the nuclear saturation density  $\rho_0$,  the pressure of pure neutron matter {for the densities 0.04, 0.08, 0.12 and 0.16~fm$^{-3}$} from N$^3$LO calculation in $\chi$EFT \cite{Hebeler2013}, taking  {2 $\times$} N$^3$LO data  uncertainty in the likelihood, and the NS maximum mass above 2.0 M$_\odot$. 
    
    \item {\it Set 1, 2, 3 and 4} include the  {\it minimal set} and, in addition, the M$_{\rm dUrca}$ pseudo constraint fixed, respectively, at 1.4, 1.6, 1.8 and 2.0 $M_\odot$. 
\end{itemize}

For sampling, we have adopted the Nested Sampling algorithm in the Bayesian Inference Library (BILBY) \cite{Ashton2019}  by invoking a {\it PyMultiNest} sampler \cite{Buchner:2014nha,buchner2021nested}. We generate samples for starting 4000 {\it"n-live"} points for each and every set separately. There are around 20000 final EOS  selected for every case by calling approx $7 \times 10^7$ model parameters. It is to be noted that in the Nested Sampling,  the posterior is broken into many nested “slices” with starting {\it"n-live"} points,  samples are generated from each of them and then recombined  to reconstruct the original distribution.

{\it Results.}
\begin{figure}
\centering
\includegraphics[width=0.42\textwidth,angle=0]{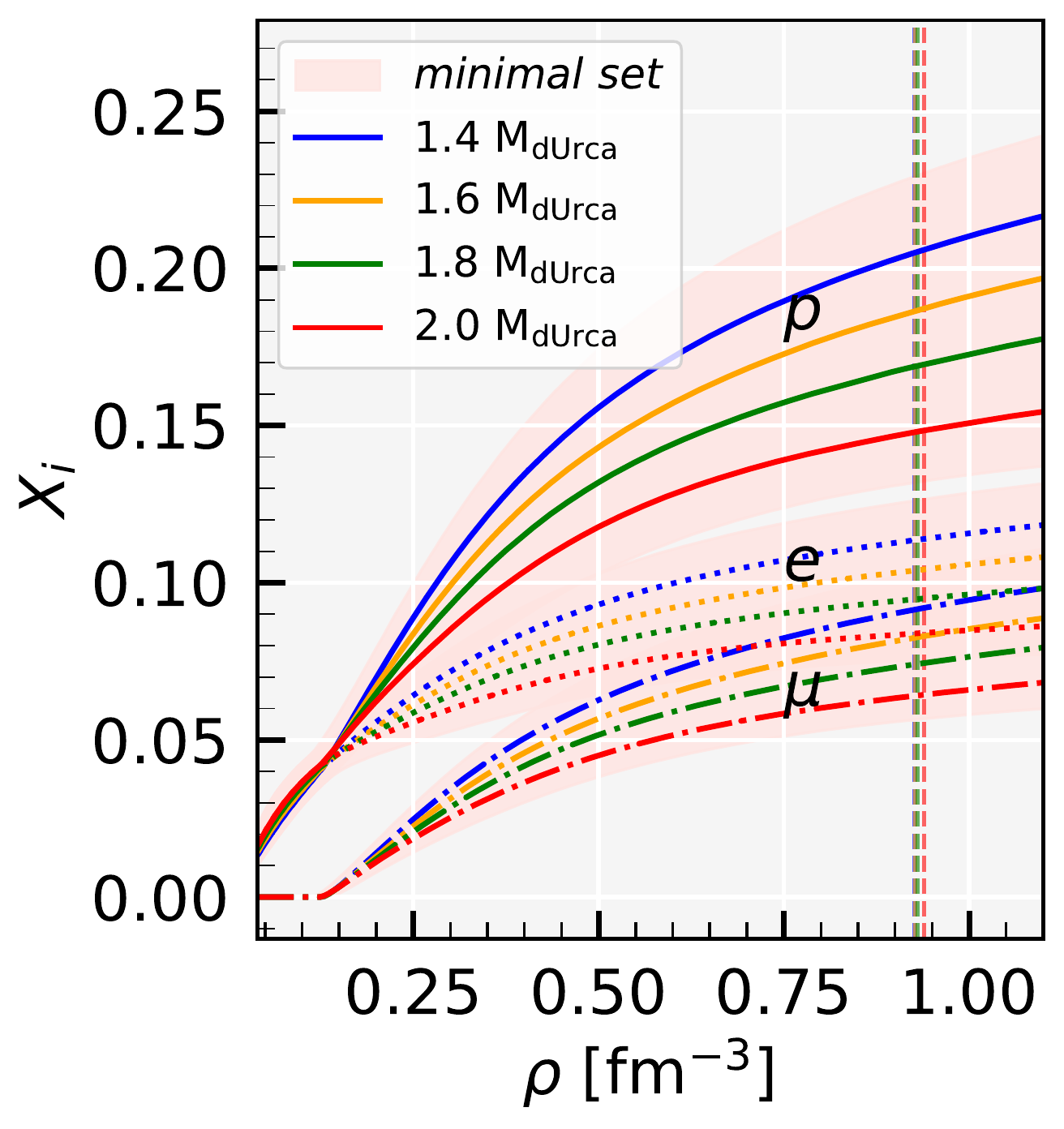}
\caption{The 90\% CI of the particle fraction $X_i$ proton (p), electron (e) and muons ($\mu$) fraction obtained for the {\it minimal set}. The median dependency of  the proton (solid), the electron (dotted) and the muon ($\mu$) fraction with the additional dUrca mass constraint  for 1.4, 1.6, 1.8 and 2.0 M$_\odot$ NS is also shown. Vertical lines indicate central densities.\label{xp}}
\end{figure}
We will first discuss  the most promising probes to constrain the symmetry energy at supra saturation densities.
We infer a large family of $\beta-$equilibrium EOSs along with its isoscalar and isovector components, i.e, the  symmetric nuclear matter (SNM) EOS and the  nuclear symmetry energy,  within a DDH framework informed by the {\it minimal set}, as mentioned above, in a Bayesian framework. With these EOSs,  we calculate the Pearson correlation coefficient at different densities between the individual  EOS components and  different star properties. In Fig. \ref{cor}, we plot the correlation coefficient $r_{x,y}$ as a function of baryon density for  $x$ =\{ $\beta-$equilibrium pressure (solid lines), the SNM pressure (dashed lines) and  the nuclear symmetry energy (dot-dashed lines)\} and  $y$=\{ the NS maximum mass M$_{\rm max}$, the minimum NS mass at which nucleonic direct Urca occurs M$_{dUrca}$ and the dimensionless tidal deformability $\Lambda_{1.4}$ for 1.4 M$_{\odot}$ NS\}.   We find that the NS maximum mass is strongly correlated with the $\beta-$equilibrium  and the SNM pressures for densities  above 0.55 fm$^{-3}$. The values of the correlation coefficients  for those cases are $\sim 0.99$ and $\sim 0.94$, respectively. However, the M$_{\rm max}$ is not correlated with the  nuclear symmetry energy at any densities: this behavior is not surprising since one expects that the  NS maximum mass mainly depends on the isoscalar part of the EOS. The dimensionless tidal deformability for  1.4 M$_\odot$ NS shows a strong correlation with NS matter EOS at $\sim 2 \times \rho_0$ density and the correlation gets weaker as density increases. It does not show any strong correlation neither with the SNM EOS nor with the symmetry energy \cite{Tsang:2020lmb}. On the other hand, the NS mass M$_{\rm dUrca}$  is only  strongly correlated  with  the  nuclear symmetry energy for densities above 0.5 fm$^{-3}$. It does not show any correlation  either with the NS matter EOS and or with the SNM EOS. We conclude that  NS properties, such as the NS maximum mass, the  radius and tidal deformability can  constrain {at best the EOS for the NS matter and symmetric nuclear matter but not the density dependence of symmetry energy which plays  a crucial role in determining the chemical composition of NS.} {In this context, the M$_{\rm dUrca}$, which is strongly correlated with the symmetry energy at supra-saturation densities, seems to be a promising probe to infer the NS composition of the core.} In the next step of the analysis, we  explore  the density dependence of nuclear symmetry energy further by invoking few additional pseudo constraints on the M$_{\rm dUrca}$  along with the {\it minimal set}.

{Since the nuclear symmetry energy shows a strong correlation with  M$_{\rm dUrca}$ for densities above 0.5 fm$^{-3}$,  it should be possible to express the density dependence of symmetry energy  $S(\rho)$ in a linear relation as 
\begin{equation} \frac{S(\rho)}{\rm MeV}= \frac{a(\rho)}{{\rm MeV}}~\frac{{\rm M}_{\rm dUrca}}{M_{\odot}} + \frac{b(\rho)}{{\rm MeV}} \label{srho}
\end{equation}
The fitted coefficients $a(\rho)$ and $b(\rho)$  are given by (i) $a(\rho)/{\rm MeV}=-56.8095~\rho/{{\rm fm}^{-3}}+4.8828$ and (ii) $b(\rho)/{\rm MeV}=166.0152~\rho/{{\rm fm}^{-3}}+30.1419$. 
For a given  M$_{\rm dUrca}$, one can calculate the symmetry energy for any density above 0.5 fm$^{-3}$ with relation (\ref{srho}).}

{Fig. \ref{esym}, left panel, shows the nuclear symmetry energy as a function of density. The colored bands for densities above 0.5 fm$^{-3}$ were  calculated with the linear relations  between the symmetry energy $S(\rho)$ and  the dUrca mass M$_{\rm dUrca}$,  for different ranges of M$_{\rm dUrca}$. The tonality of color varies from red to sky blue for M$_{\rm dUrca}$  ranging from 1.2 to 2.4 M$_\odot$ in steps of 0.2M$_\odot$. 
We also plot the median values of the  symmetry energy obtained for marginalized posterior distribution of each of the sets 1 to 4 defined above. These median values of the symmetry energy under the condition that the dUrca mass is 1.4, 1.6, 1.8 and 2.0 M$_\odot$ match very well with the calculated symmetry energy {using Eq. (\ref{srho})}. We also compare the symmetry energy with the nuclear structure studies involving excitation energies of isobaric analog states (IASs) \cite{Danielewicz:2013upa}. For all the four cases, for sets 1 to 4 corresponding to dUrca for  NS mass 1.4 - 2.0 M$_\odot$, the symmetry energy is in good agreements with IAS data.}

{We also found that the slope of the symmetry energy $L$ ($L(\rho)=3~\rho_0~ \partial S(\rho)/\partial \rho$) is strongly anti correlated with M$_{\rm dUrca}$ at 0.375 fm$^{-3}$ ($\sim 2.5~\rho_0$) baryon density and the Pearson correlation coefficient is 0.9. The linear relation between $L$ and M$_{\rm dUrca}$ for 0.375 fm$^{-3}$ density is as follows,
\begin{equation} \frac{L}{\rm MeV}= \frac{-31.224}{{\rm MeV}}~\frac{{\rm M}_{\rm dUrca}}{M_{\odot}} + \frac{104.339}{{\rm MeV}} \label{lrho}
\end{equation}
For nucleonic dUrca processes to occur inside NS with masses between 1.6-1.8 M$_\odot$ the value of the slope of the symmetry energy $L$ at density $\sim 2.5 ~ \rho_0$ is found to be in the range 54- 48 MeV.} 

In the right panel of Fig. \ref{esym}, the low-density  PNM pressure obtained with our EOS sets is compared with the results of  the  $\chi$EFT (N$^{3}$LO) calculation  \cite{Hebeler2013}. The   90\% CI of low-density  PNM pressure corresponding  to the {\it minimal set}  overlaps partially  with the 2$\sigma$  $\chi$EFT uncertainty band. It is to be noted that PNM pressure obtained for the set at which dUrca happens at 2.0 M$_\odot$ is within the  $\chi$EFT uncertainty band at 1$\sigma$. As the NS mass for dUrca onset decreases the PNM pressure gets harder, and the pressure for the EOS that predicts  M$_{\rm dUrca}$ at  1.4 M$_\odot$, lies just above the 2$\sigma$ $\chi$EFT uncertainty band. {The constraints from the PNM EOS seem to rule out the values of M$_{\rm dUrca} \lesssim 1.4 M_\odot$, which is in line with  investigations based on the analysis of  NS cooling curves that suggest  M$_{\rm dUrca}$ $\sim 1.6 - 1.8$ M$_\odot$ \cite{Beznogov:2015ewa}.}

{The proton, electron and muon fractions are plotted as a function of the baryonic density in Fig. \ref{xp}. The 90\% CI of particle fraction  obtained for the posterior distributions of the DDH parameters corresponding to the {\it minimal set} fit data is plotted in light pink band. We also plot the median particle fraction for each individual case corresponding to the Sets 1 to 4  obtained imposing that the nucleonic dUrca occurs for NS masses 1.4-2.0 M$_\odot$ in addition to {\it minimal set} fit data. The NS central densities for all the cases lie below 1 fm$^{-3}$, and are represented by thin vertical lines with the color matching the set color.}
{The EOS corresponding to the four medians in Fig. \ref{xp} predict similar properties for large mass stars, and only differ when considering low mass stars: stars with $M_{dUrca}=1.4M_\odot$ have a $\sim$2\% larger radius and $\sim 10\%$ larger tidal deformability compared with  stars with $M_{dUrca}=1.8M_\odot$ (see Table II in the supplementary material). This is in line with conclusions drawn in several works where it was shown that low mass stars are sensitive to the symmetry energy behavior \cite{Alam:2016cli, Malik:2018zcf, Ferreira:2019bny} or with neutron skin in $^{208}$Pb
\cite{Carriere:2002bx}.}

{\it Conclusions.}
Within a Bayesian inference approach applied to a RMF model with density dependent couplings, the parameters of the model were constrained using a fit {\it minimal set}. {Since DDH RMF models  do not predict the occurrence of nucleonic dUrca inside NS, because their $\rho$-meson coupling tends to zero at high densities,}
a parameter {was introduced explicitly to}  control the symmetry energy at high densities. In a second step, also an hypothetical dUrca mass  corresponding to the mass of the star where the nucleonic direct Urca process sets in, was imposed.  It was possible to  show that: (i) above the baryonic density 0.5fm$^{-3}$ the symmetry energy is very well correlated with the M$_{\rm dUrca}$, and the NS maximum mass M$_{\rm max}$ is very well correlated with the $\beta$-equilibrium and SNM pressure; (ii) from the correlation between the symmetry energy and M$_{\rm dUrca}$, a linear relation between these two quantities was derived; (iii)  applying the deduced correlation the density dependence of symmetry energy was predicted taking into account the possible value of M$_{\rm dUrca}$; (iv) this dependence was confirmed applying a second Bayesian inference that also imposes  the conditions on  M$_{\rm dUrca}$. 

It was, therefore, shown that the M$_{\rm dUrca}$ mass may be considered a strong constraint to determine the high density dependence of the symmetry energy. The 90\% CIs for nuclear symmetry energy $S(\rho)$ at $\rho=$0.56 fm$^{-3}$ are $\in[74,82]$ MeV and $\in[69,76]$ MeV for  1.6 M$_{\rm dUrca}$ and 1.8 M$_{\rm dUrca}$, respectively. We should, however, point out that there are other physical factors the affect the dUrca cooling besides the proton fraction and that need to be estimated before M$_{\rm dUrca}$ can be taken as high density constraint for the symmetry energy: (i)  the proton and neutron pairing in the outer core \cite{Yakovlev:2000jp,Beznogov:2015ewa}; (ii) the hyperonic dUrca processes after the nucleation of hyperons inside the star \cite{Fortin:2020qin,Fortin:2021umb}. After the nucleation of hyperons also the hyperon pairing affects the  cooling rates \cite{Raduta:2017wpp,Raduta:2019rsk}. If observations allow the estimation of the M$_{\rm dUrca}$ the corresponding symmetry energy obtained through the Eq. (\ref{esym}) will be a lower bound, since pairing may shift the dUrca to larger densities, depending on the central densities and the pairing gaps. The opening of hyperonic channels will lower M$_{\rm dUrca}$ if pairing is not considered, and in this case Eq. (\ref{esym}) would predict an upper bound for the  symmetry energy. However, hyperon pairing will turn the analysis more complicated. {The inclusion of hyperons will be considered in a future study}.

\medskip
\begin{acknowledgments}
This work was partially supported by national funds from FCT (Fundação para a Ciência e a Tecnologia, I.P, Portugal) under the Projects No. UID/\-FIS/\-04564/\-2019, No. UIDP/\-04564/\-2020, No. UIDB/\-04564/\-2020, and No. POCI-01-0145-FEDER-029912 with financial support from Science, Technology and Innovation, in its FEDER component, and by the FCT/MCTES budget through national funds (OE). The authors acknowledge the Laboratory for Advanced Computing at University of Coimbra for providing {HPC} resources that have contributed to the research results reported within this paper, URL: \hyperlink{https://www.uc.pt/lca}{https://www.uc.pt/lca}.
\end{acknowledgments}

%
\newpage

\onecolumngrid\
\clearpage
\setcounter{figure}{0}
\section{Supplemental Material}

\begin{table*}[h!]
\caption{The maximum Pearson correlation coefficient obtained for {\it minimal sets} at densities $\rho$ between the individual  EOS components and  different star properties. The maximum correlation coefficient $r_{x,y}$ at a given $\rho$ in fm$^{-3}$ for $x$ =\{ $\beta-$equilibrium pressure ($P_\beta$), the SNM pressure ($P_{\rm SNM}$), the nuclear symmetry energy ($S(\rho)$) and slope of the symmetry energy $L(\rho)$ \} and $y$=\{ the NS maximum mass M$_{\rm max}$, the radius $R_{1.4}$ and $R_{2.075}$ for 1.4 and 2.075 $M_\odot$ NS, respectively, the dimensionless tidal deformability $\Lambda_{1.4}$ for 1.4 M$_{\odot}$ NS and the minimum NS mass at which nucleonic direct Urca occurs M$_{\rm dUrca}$ \}.}
\setlength{\tabcolsep}{10.5pt}
      \renewcommand{\arraystretch}{1.4}
\begin{tabular}{ccccccccccc}
\hline \hline
\multirow{3}{*}{x} & \multicolumn{10}{c}{y}                                                                                                                                                          \\
                   & \multicolumn{2}{c}{$M_{\rm max}$} & \multicolumn{2}{c}{$R_{1.4}$} & \multicolumn{2}{c}{$R_{2.075}$} & \multicolumn{2}{c}{$\Lambda_{1.4}$} & \multicolumn{2}{c}{$M_{\rm dUrca}$} \\
                   & $\rho$         & $r_{x,y}$        & $\rho$       & $r_{x,y}$      & $\rho$        & $r_{x,y}$       & $\rho$          & $r_{x,y}$         & $\rho$          & $r_{x,y}$         \\ \hline
$P_{\beta}$        & 0.662          & 0.99             & 0.255        & 0.96           & 0.710         & 0.64            & 0.279           & 0.99              & 0.136           & -0.74             \\
$P_{\rm SNM}$      & 0.710          & 0.96             & 0.303        & 0.74           & 0.710         & 0.62            & 0.327           & 0.85              & \multicolumn{2}{c}{NA}              \\
$S(\rho)$          & \multicolumn{2}{c}{NA}            & 0.303        & 0.73           & \multicolumn{2}{c}{NA}          & 0.327           & 0.54              & 0.560           & -0.92             \\
$L(\rho)$          & \multicolumn{2}{c}{NA}            & 0.231        & 0.67           & \multicolumn{2}{c}{NA}          & 0.231           & 0.54              & 0.375           & -0.90             \\ \hline
\end{tabular}
\end{table*}

\begin{figure*}
\centering
\includegraphics[width=0.9\textwidth,angle=0]{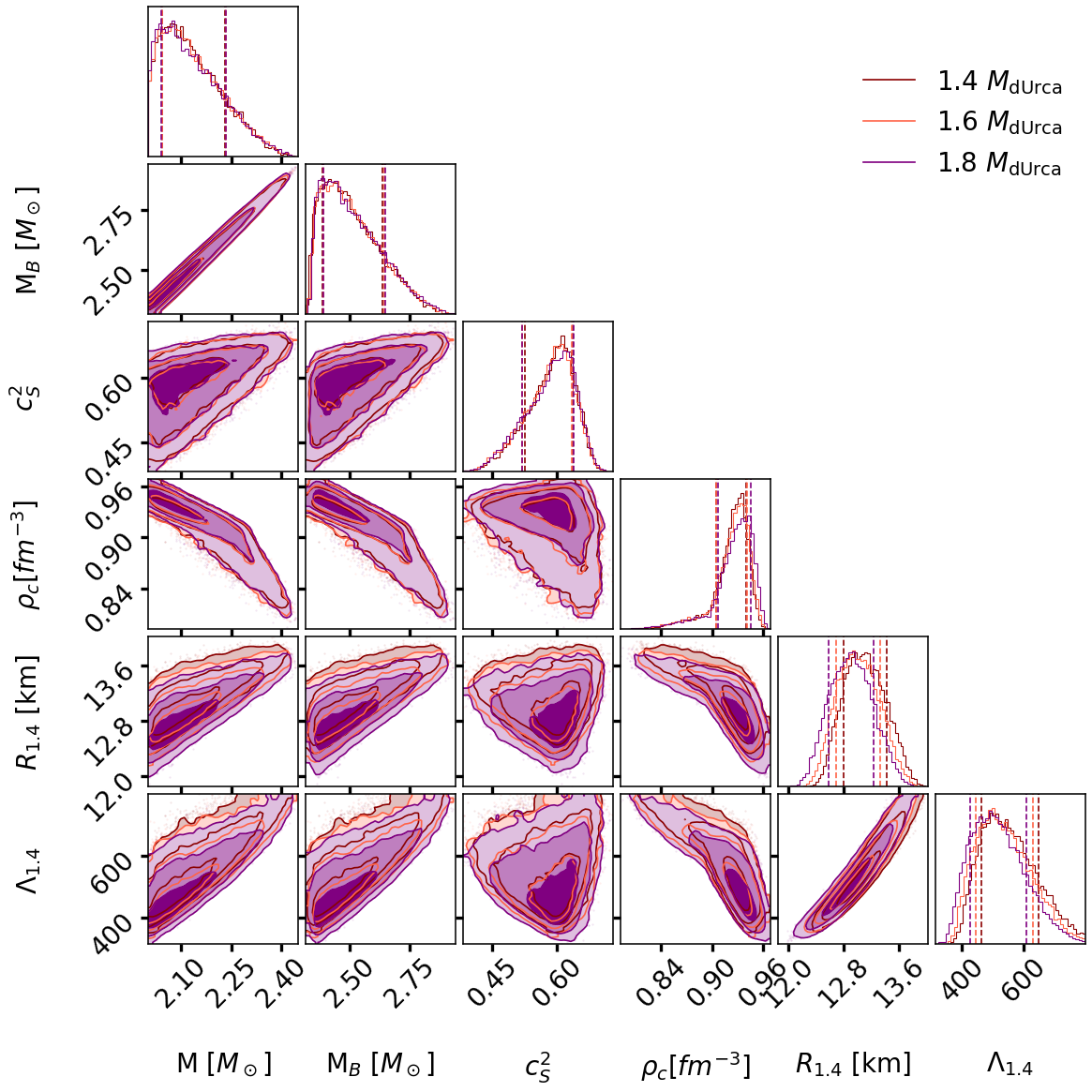}
\caption{Corner plots for the marginalized posterior distributions (PDs) of neutron star properties, namely gravitational mass $M_{\rm max}$, baryonic mass $M_{\rm B, max}$, the square of central speed-of-sound $c_s^2$, the central baryonic density $\rho_{c}$, the radius $R_{1.4}$ and  the dimensionless tidal deformability $\Lambda_{1.4}$ for 1.4 $M_\odot$ NS
for the model with 1.4 $M_{\rm dUrca}$ (dark red), 1.6 $M_{\rm dUrca}$ (salmon) and 1.8 $M_{\rm dUrca}$ (violet) constraints considered in Set 1, 2 and 3, respectively. The vertical lines indicate 68\% min, median and 68\% max CI, respectively, and the different tonalities from dark to light indicate, respectively, the 1$\sigma$, 2$\sigma$, and 3$\sigma$ CI.}
\end{figure*}

\begin{table*}[]
\caption{The median values and the associated  90\% CI of the NS properties the gravitational mass $M_{\rm max}$, baryonic mass  $M_{\rm B, max}$, radius $R_{\rm max}$, central energy density $\varepsilon_c$, central number density for baryon $\rho_c$ and square of central speed-of-sound $c_s^2$ of the maximum mass NS, as well as the radius $R_M$  and  the dimensionless tidal deformability $\Lambda_M$ for a NS having a solar mass $M$. {The $\tilde \Lambda_{q=1}$ refers to GW170817 and corresponds to $M=1.36\,M_{\odot}$.}} 
\setlength{\tabcolsep}{8.5pt}
      \renewcommand{\arraystretch}{1.4}
\begin{tabular}{ccccccccccc}
\hline \hline 
\multirow{3}{*}{Quantity} & \multirow{3}{*}{Units} & \multicolumn{3}{c}{$M_{\rm dUrca}=1.4 ~M_\odot$}        & \multicolumn{3}{c}{$M_{\rm dUrca}=1.6 ~M_\odot$}        & \multicolumn{3}{c}{$M_{\rm dUrca}=1.8 ~M_\odot$}        \\ \cline{3-11} 
                          &                        & \multirow{2}{*}{median}   & \multicolumn{2}{c}{90\% CI} & \multirow{2}{*}{median}   & \multicolumn{2}{c}{90\% CI} & \multirow{2}{*}{median}   & \multicolumn{2}{c}{90\% CI} \\
                          &                        &                           & min           & max         &                           & min           & max         &                           & min           & max         \\ \hline
$M_{\rm max}$             & M $_\odot$             & $2.119$                   & $2.017$       & $2.304$     & $2.120$                   & $2.017$       & $2.310$     & $2.118$                   & $2.015$       & $2.309$     \\
$M_{\rm B, max}$          & M $_\odot$             & $2.487$                   & $2.350$       & $2.735$     & $2.492$                   & $2.354$       & $2.746$     & $2.492$                   & $2.357$       & $2.748$     \\
$c_{s}^2$                 & $c^2$                  & $0.59$                    & $0.48$        & $0.66$      & $0.59$                    & $0.47$        & $0.66$      & $0.59$                    & $0.47$        & $0.66$      \\
$\rho_c$                  & fm$^{-3}$              & $0.927$                   & $0.872$       & $0.947$     & $0.927$                   & $0.868$       & $0.949$     & $0.930$                   & $0.870$       & $0.953$     \\
$\varepsilon_{c}$         & MeV fm$^{-3}$          & $1173$                    & $1130$        & $1173$      & $1173$                    & $1119$        & $1173$      & $1173$                    & $1130$        & $1173$      \\
$R_{\rm max}$             & \multirow{8}{*}{km}    & $11.30$                   & $10.83$       & $11.86$     & $11.28$                   & $10.80$       & $11.85$     & $11.24$                   & $10.74$       & $11.83$     \\ \\ 
$R_{0.8}$                 &                        & $13.49$                   & $13.08$       & $13.91$     & $13.33$                   & $12.91$       & $13.74$     & $13.18$                   & $12.71$       & $13.61$     \\
$R_{1.0}$                 &                        & $13.33$                   & $12.92$       & $13.76$     & $13.19$                   & $12.77$       & $13.61$     & $13.06$                   & $12.60$       & $13.50$     \\
$R_{1.2}$                 &                        & $13.21$                   & $12.78$       & $13.68$     & $13.10$                   & $12.66$       & $13.55$     & $12.98$                   & $12.51$       & $13.46$     \\
$R_{1.4}$                 &                        & $13.09$                   & $12.63$       & $13.61$     & $13.00$                   & $12.53$       & $13.51$     & $12.90$                   & $12.41$       & $13.43$     \\
$R_{1.6}$                 &                        & $12.93$                   & $12.42$       & $13.52$     & $12.85$                   & $12.34$       & $13.44$     & $12.77$                   & $12.24$       & $13.37$     \\
$R_{1.8}$                 &                        & $12.69$                   & $12.10$       & $13.38$     & $12.63$                   & $12.04$       & $13.32$     & $12.56$                   & $11.95$       & $13.26$     \\
$R_{2.075}$               &                        & $12.21$                   & $11.37$       & $13.09$     & $12.18$                   & $11.34$       & $13.07$     & $12.14$                   & $11.28$       & $13.04$     \\ \\ 
$\Lambda_{0.8}$           & \multirow{8}{*}{-}     & $12044$                   & $9994$        & $14515$     & $11384$                   & $9376$        & $13741$     & $10777$                   & $8789$        & $13130$     \\
$\Lambda_{1.0}$           &                        & $3866$                    & $3171$        & $4772$      & $3693$                    & $3012$        & $4556$      & $3520$                    & $2840$        & $4382$      \\
$\Lambda_{1.2}$           &                        & $1335$                    & $1074$        & $1701$      & $1287$                    & $1029$        & $1638$      & $1237$                    & $978$         & $1586$      \\
$\Lambda_{1.4}$           &                        & $543$                     & $425$         & $717$       & $527$                     & $409$         & $696$       & $509$                     & $391$         & $678$       \\
$\Lambda_{1.6}$           &                        & $214$                     & $160$         & $299$       & $209$                     & $155$         & $293$       & $203$                     & $149$         & $287$       \\
$\Lambda_{1.8}$           &                        & $88$                      & $61$          & $135$       & $87$                      & $60$          & $133$       & $85$                      & $58$          & $131$       \\
$\Lambda_{2.075}$         &                        & $24$                      & $12$          & $44$        & $23$                      & $12$          & $44$        & $23$                      & $12$          & $44$        \\
$\tilde \Lambda_{q=1}$    &                        & $665$                     & $524$         & $870$       & $645$                     & $504$         & $843$       & $622$                     & $481$         & $820$       \\ \hline
\end{tabular}
\end{table*}

\end{document}